\documentclass[useAMS,usenatbib]{mn2e}

\usepackage{graphicx}
\usepackage{epstopdf}

\title[The gravitational wave `probability event horizon' for
   double neutron star mergers]{The gravitational wave `probability event horizon' for
   double neutron star mergers}
\author[D. M. Coward, M. Lilley, E. J. Howell, R. R. Burman and D. G. Blair]{D. M. Coward,\thanks{E-mail:
coward@physics.uwa.edu.au} M. Lilley, E. J. Howell, R. R. Burman
and D. G. Blair \\School of Physics, University of Western
Australia, M013, Crawley WA 6009, Australia}
\begin{document}

\date{\today}

\pagerange{\pageref{firstpage}--\pageref{lastpage}} \pubyear{2005}

\maketitle

\label{firstpage}

\begin{abstract}
Gravitational waves generated by the final merger of double
neutron star (DNS) binary systems are a key target for the
gravitational wave (GW) interferometric detectors, such as LIGO,
and the next generation detectors, Advanced LIGO. The cumulative GW
signal from DNS mergers in interferometric data will manifest as
`geometrical noise': a non-continuous stochastic background with a unique statistical signature dominated by the spatial and temporal distribution of the sources. Because geometrical noise is highly non-Gaussian, it could potentially be used to identify the presence of a stochastic GW background from DNS mergers. We demonstrate this by fitting to a simulated distribution of  transients  using a model for the DNS merger rate and idealized Gaussian detector noise. Using the cosmological `probability event horizon' concept and recent bounds for the Galactic DNS merger rate, we calculate the evolution of the detectability of DNS mergers with observation time. For Advanced LIGO sensitivities and a detection threshold assuming optimal filtering, there is a 95\% probability that a minimum of one DNS merger signal will be detectable from the ensemble of events comprising the stochastic background during 12--211 days of observation. For initial LIGO sensitivities, we identify an interesting regime where there is a 95\% probability that at least one DNS merger with signal-to-noise ratio $>$ unity will occur during 4--68 days of observation. We propose that there exists an intermediate detection regime with pre-filtered signal-noise-ratio less than unity, where the DNS merger rate is high enough that the geometrical signature could be identified in interferometer data.
\end{abstract}


\section{Introduction}

Three long-baseline laser interferometer GW detectors have been,
or are nearly, constructed. The US LIGO (Laser Interferometer
Gravitational-wave Observatory) has started observation with two
4-km arm detectors situated at Hanford, Washington, and
Livingston, Louisiana; the Hanford detector also contains a 2-km
interferometer. The Italian/French VIRGO project is commissioning
a 3-km baseline instrument at Cascina, near Pisa. There are
detectors being developed at Hannover (the German/British GEO
project with a 600-m baseline, which had its first test runs in
2002) and near Perth (the Australian International Gravitational
Observatory, AIGO, initially with an 80-m baseline). A detector at
Tokyo (TAMA, 300-m baseline) has been in operation since 2001. The
astrophysical detection rates are expected to be low for the
current interferometers, such as `Initial LIGO', but
second-generation observatories with high optical power are in the
early stages of development; these `Advanced' interferometers have
target sensitivities that are predicted to provide a practical
detection rate.

Double neutron star (DNS) binary mergers are potentially among the
strongest GW sources and are a key search target for
interferometric observatories \citep{Thorne02}. Eight close DNS
binary systems are known to exist in the Galaxy as a subset of the
observed radio binary pulsar population \citep{Hulse75,Wol91}.
Energy loss from GW emission \citep{Tayl89} causes an orbital
in-spiral until the binary system coalesces, resulting in a burst
of GWs usually described as a `chirp' signal. The discovery of the
DNS binary systems containing PSRs J0737--3039 \citep{Burg03} and
J1756$-$2251 \citep{faulk05} using the Parkes radio telescope
brings the number of known DNS systems in the Galactic disk to
merge within the Hubble time to four; this excludes the PSR
B2127+11C system as it is in a globular cluster and probably not
formed by binary evolution (see Table 1). With an orbital period
of only 2.45 hr, the double pulsar system J0737--3039A,B will
coalesce in only 87 Myr, a factor of 3.5 shorter than the
coalescence time of the Hulse-Taylor PSR B1913+16 system, which
yielded the first evidence for gravitational radiation. Predicted
merger rates are dominated by the  J0737--3039 system, because of
its proximity, short coalescence time and the difficulty of its
discovery \citep{faulk05}. In contrast, PSR J1756--2251's
parameters are similar to those of previously known pulsars; thus,
its very recent addition isn't expected to significantly alter the
predicted merger rate.

\begin{table*}
\centering
\begin{minipage}{130mm}
\caption{DNS systems containing radio pulsars arranged in order of
increasing estimated coalescence time. The first entry is the
double pulsar system; the second system is in a globular cluster
(M 15); the third is the Hulse-Taylor binary. Here, $e$ is the
orbital eccentricity, $\tau_{\rm c}$ is the pulsar characteristic
age and $\tau_{\rm GW}$ is the time remaining to coalescence due
to emission of gravitational radiation. The relationship between a
pulsar's characteristic and true ages is not clear, but we suppose
the total coalescence time from birth to be $\tau_{\rm
c}+\tau_{\rm GW}$. Bold type indicates DNS systems that will
coalesce in less than $10^{10}$ yr; the gobular-cluster system B2127+11C is excluded because its formation history is
probably different from those of the Galactic-disk systems: it is
likely that this system has undergone partner exchange from close
interactions. References: (1) Lyne et al. (2004); (2) Burgay et
al. (2003); (3) Anderson et al. (1990); (4) Hulse \& Taylor
(1975); (5) Faulkner et al. (2005); (6) Wolszczan (1990); (7) Lyne
et al. (2000); (8) Nice et al. (1996); (9) Champion et al. (2004).
Adapted from Faulkner et al. (2005)}
\begin{tabular}{|l|l|l|l|l|l|l|l|}\hline
PSR & pulsar & orbital
 & $e$ & system &$\tau_{\rm c}$&
$\tau_{\rm GW}$ & Ref.\\
& period  & period  &   & mass   &   &  &\\
    & (ms)  & (hr)  &   & (M$_{\odot})$&(Myr)& (Myr)&\\
    \hline\\
    {\bf J0737$-$3039A,B}   & 23, 2773  & 2.45  & 0.088 & 2.58 &
210,50 & 87 & (1,2)
\\
B2127+11C   & 31 & 8.04  & 0.681 & 2.71  & 969   & 220 & (3)
   \\
{\bf B1913+16}    & 59 & 7.75  & 0.617 & 2.83  & 108   & 310 & (4)
\\
{\bf J1756$-$2251}    & 28 & 7.67  & 0.181 & 2.57  & 444   & 1690
& (5)
 \\
{\bf B1534+12}    & 38 & 10.10 & 0.274 & 2.75  & 248   & 2690 &
(6)
    \\
J1811$-$1736  & 104  & 449 & 0.828 &  $2.6$ &  900 &
$1.7\times 10^6$ & (7)\\
J1518+4904  & 41  & 207  & 0.249 & 2.61  &  16,000 & $2.4\times
10^6$ & (8)
\\
J1829$+$2456  & 41  & 28.2 & 0.139 &  2.5 & --& $6\times10^4$& (9) \\
\hline
\end{tabular}
\end{minipage}
\end{table*}

In order to estimate the DNS coalescence rate, Kalogera et al.
(2001) used a semi-empirical approach, based on the physical
properties of known DNS systems and pulsar survey selection
effects, to obtain scale factors that correct for the unobserved
fraction of existing systems. In an alternative approach, Kim et
al. (2004) generated the probability distribution of merger rates
using Monte Carlo simulations in which they populate a model
galaxy with a specified number of NS binaries whose properties are
those of known systems; they also model survey selection effects to
deduce detection rates and their statistical significance. With
this model and more up-to-date pulsar survey results that include
J0737--3039, \cite{Kalog04} presented updated bounds for the
coalescence rate for Galactic disk DNS systems, yielding
$R_{\mathrm {DNS}} = 83^{+209.1}_{-66.1}$ Myr$^{-1}$, and derived
detection rates for Initial and Advanced LIGO of
$34.8^{+87.6}_{-27.7}\times 10^{-3}$ yr$^{-1}$ and
$168.8^{+470.5}_{-148.7}\times$ yr$^{-1}$ respectively,
factors 5--7 higher than previous estimates.

Others \citep{Piran02,Ando04} assume a simple parametrization of
merger rates with merger probability $P_m(t) \propto t^\alpha$,
where $t$ is the time from formation of the binary and $\alpha$ is
a constant; the distribution of mergers as a function of time is
simply given by the convolution of the star formation rate with
the distribution $P_m(t)$. 
\cite{Reg05}, using numerical simulations, derive a probability distribution for DNS merger time,
 $\tau$, of the form $P(\tau)=0.087/ \tau$, and find that mergers are possible
between $2 \times 10^5$ yr and the age of the Universe.
Using both evolutionary and statistical models, \cite{Reg05} find a
Galactic merger rate of $1.7 \times 10^{-5}$ yr$^{-1}$, 
similar to the lower bound  calculated by \cite{Kalog04}.

A purely theoretical approach to
estimate coalescence rates can be based on the predictions of
population synthesis models that attempt to simulate the entire
pulsar population \citep{Lip97,Bel02}. However, the size of the
parameter space and associated uncertainties involved seriously
limit this way of inferring merger rates.  For instance, new
theoretical models of merger formation can have considerable
impact: Recent simulations by Chaurasia and Bailes (2005) show
that significant kick velocities imparted to neutron stars at
birth give rise to highly eccentric orbits resulting in
accelerated orbital decay and greatly reduced lifetimes. Their
work indicates that a non-negligible fraction of DNS systems could
merge within times as short as 10 yrs. This implies that there
could be a selection effect favouring the observation of those
binary pulsar systems that coalesce over relatively long times.

\begin{figure}
\includegraphics[scale=0.6]{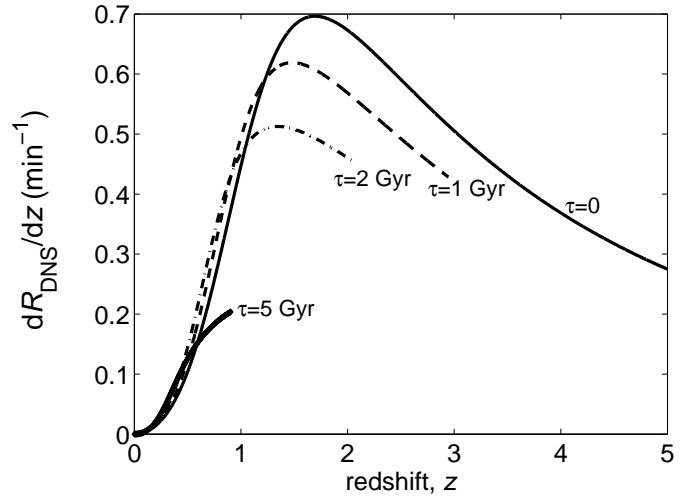}
\caption{The differential DNS merger rate as a function of
redshift $z$ using the lower Galactic rate of 17 Myr$^{-1}$ from
Kalogera et al. (2004) and four merger times, $\tau =$ (0,1,2,5)
Gyr, assumed the same for all systems. The curves are calculated
using the SFR evolution factor obtained from the parametrized
model labelled SF2 in Porciani \& Madau (2001) and a
flat-$\Lambda$ cosmology with $\Omega_{\mathrm m}=0.3$ and
$\Omega_{\mathrm \Lambda}=0.7$, assuming a SFR cut-off at $z=5$;
the cut-off projects forward in time to the end points of the
time-delayed curves. A time delay of 1--5 Gyr has minimal effect
on $\mathrm{d}R_{\mathrm{DNS}}/\mathrm{d}z$ at small $z$ compared
to the differential rate locked to the evolving
SFR.}\label{frates}
\end{figure}

Our interest is in the detection of a GW background generated by
DNS merger events throughout the cosmos. As a starting point, we
adopt the Galactic coalescence rates of \cite{Kalog04} as a basis
for estimating the rate density in our local intergalactic
neighbourhood. To project this throughout the Universe, we assume
that the DNS system formation rate follows the evolving star
formation rate (SFR), for DNS systems with relatively short merger times
of tens to hundreds of millions of years. In this way, we employ
the updated Galactic DNS merger rate estimates as input to a model
that describes the observer's evolving record of DNS mergers in
terms of a `probability event horizon' (PEH). We describe how
the detectability of DNS mergers evolves with observation time.

We use the PEH concept \citep{cow05} to describe the cumulative GW
signal from cosmic DNS mergers and show that the signal is
dominated by the spatial and temporal distribution of events.
Instead of describing such a signal as `stochastic', we argue that
the signal, although temporally random, has a unique signature. We
use the term `geometrical noise' to distinguish it from the
stationary noise usually associated with a primordial GW
stochastic background.

\section{The cosmological probability event horizon}
In the standard Friedmann cosmologies, one can express the
differential DNS merger rate as the merger rate in the redshift
shell $z$ to $z+{\mathrm d}z$:
\begin{equation}\label{drdz}
\mathrm{d}R_{\mathrm{DNS}} =
\frac{\mathrm{d}V}{\mathrm{d}z}\frac{r_0 e(z)}{1+z} \mathrm{d}z
\,,
\end{equation}
\noindent where $\mathrm{d}V$ is the cosmology-dependent co-moving
volume element and  $R_{\mathrm{DNS}}(z)$ is the all-sky ($4\pi$
solid angle) DNS merger rate, as observed in our local frame, for
sources out to redshift $z$. Source rate density evolution is
accounted for by the dimensionless evolution factor $e(z)$; this
is normalized to unity in the present-epoch universe $(z=0)$ so
$r_0$ is the $z=0$ rate density. The Galactic DNS merger rate is
converted to a rate per unit volume, $r_0$, using the conversion
factor $10^{-2}$ from \cite{Ando04} for the number density of
galaxies in units of Mpc$^{-3}$. The $(1 + z)$ factor accounts for
the time dilation of the observed rate by cosmic expansion,
converting a source-count equation to an event rate equation.

We assume a `flat-$\Lambda$' cosmology, in which the density
parameters of matter and vacuum energy sum to unity; we use
$\Omega_{\mathrm m}=0.3$ and $\Omega_{\mathrm \Lambda}=0.7$ for
their present-epoch values and take $H_{0}=70$ km s$^{-1}$
Mpc$^{-1}$ for the Hubble parameter at $z=0$. With the assumption
that the DNS system formation rate follows the evolving star
formation rate, $e(z)$ is obtained by normalizing a SFR evolution model to
its $z=0$ value. Presently there is no consensus among astronomers
on a model of star formation history at high $z$, so we shall use
the parametrized model labelled SF2 of \cite{Porc01} scaled to the
flat-$\Lambda$ cosmology; see \cite{cow05} for details. The
cumulative DNS merger rate $R_{\mathrm{DNS}}(z)$ is calculated by
integrating the differential rate from the present epoch to
redshift $z$:
\begin{equation}\label{rat1}
R_{\mathrm{DNS}}(z)=\int_{0}^{z}(\mathrm{d}R_{\mathrm{DNS}}/\mathrm{d}z)\;\mathrm{d}z\,.
\end{equation}

Whereas core-collapse supernova events (because of their
short-lived progenitors) closely follow the evolving star
formation rate, DNS merger events are generally significantly
delayed with respect to star formation. In order to test the
sensitivity of $\mathrm{d}R_{\mathrm{DNS}}/\mathrm{d}z$ to this
delay, we have calculated it under the simplified assumption that
all such events are delayed by the same merger time $\tau$. Figure
\ref{frates} plots $\mathrm{d}R_{\mathrm{DNS}}/\mathrm{d}z$ for
four choices of $\tau$, namely 0, 1, 2 and 5 Gyr. The three
delayed curves were obtained by converting $e(z)$ to a function of
cosmic time (or, equivalently, look-back time), shifting it by
$\tau$ and then converting back to a function of $z$. Cosmic time
and $z$ are related by the evolving Hubble parameter $H(z)$, which
can be expressed in terms of the contributions of matter and
vacuum energy \citep{peebles93}:
\begin{equation}\label{hz}
h(z)\equiv H(z)/H_0 = \big[\Omega_{\mathrm m} (1+z)^3+
\Omega_{\mathrm \Lambda} \big]^{1/2}
\end{equation}
for a flat-$\Lambda$ cosmology, and cosmic time is expressed by
the formula:
\begin{equation}\label{time1}
T_{\mathrm{cos}}(z) = \int_{0}^z [(1+z)h(z)]^{-1}\mathrm{d}z\,.
\end{equation}

Figure 1 shows that a 1--2 Gyr merger time does not significantly
alter $\mathrm{d}R_{\mathrm{DNS}}/\mathrm{d}z$, especially for
small $z$. Increasing $\tau$ has the effect of reducing the
available volume for observing DNS mergers and, for $\tau>5$ Gyr,
coalescences are restricted to the volume bounded by $z\approx1$.
So, for systems such as those used by \cite{Kalog04} in their rate
estimates, we can safely assume that they quite closely follow the
evolving star formation rate. For definiteness, in the following
calculations, we assume that the DNS merger rate approximately
follows the SFR evolution factor $e(z)$.

As the DNS mergers throughout the Universe are independent of each
other, their distribution is a Poisson process in time: the
probability $\epsilon$ for at least one event to occur in a volume
out to redshift $z$ during observation time $T$ at a mean rate
$R_{\mathrm{DNS}}(z)$ is given by an exponential distribution:
\begin{equation}\label{prob2}
p(n\ge1;R_{\mathrm{DNS}}(z),T) = 1 - e^{-R_{\mathrm{DNS}}(z)
T}=\epsilon\,,
\end{equation}
with mean number of events $N_{\epsilon}\equiv
R_{\mathrm{DNS}}(z)T=\vert \mathrm{ln}(1-\epsilon)\vert$. The PEH
is defined by the minimum distance, or redshift
$z_{\epsilon}^{\mathrm{PEH}}$, for at least one event to occur
over some observation time $T$, with probability above some
selected threshold $\epsilon$. We find
$z_{\epsilon}^{\mathrm{PEH}}(T)$ by fixing $\epsilon$ and solving
the above condition numerically, thus defining the PEH.  In
practice, we shall take $\epsilon=0.95$, corresponding to a 95\%
probability of observing at least one event within $z$, so
$N_{\epsilon}=3$ is the mean number of events. Converting
$z_{\epsilon}^{\mathrm{PEH}}(T)$ to luminosity distance
$d_{\mathrm{L}\epsilon}^{\mathrm{PEH}}(T)$, and differentiating
with respect to $T$ then yields $v_{\epsilon}^{\mathrm{PEH}}(T)$:
the PEH velocity, which describes the rate at which observations
penetrate into the low-probability `tail' of the event
distribution.

\begin{table}\label{tbl-2}
\caption{The probability for at least one DNS merger to occur
within the volume bounded by the luminosity distance
$d_{\mathrm{L}}=20$ Mpc for one year of observation (Initial
LIGO) and by $d_{\mathrm{L}}=200$ Mpc for one day (Advanced LIGO)
assuming the Galactic DNS merger rate limits, $R_{\mathrm {DNS}}$,
from Kalogera et al. (2004).}
\begin{tabular}{|l|l|l|l|}\hline
 Galactic $R_{\mathrm {DNS}}$& $p(d_{\mathrm{L}}=20$ Mpc; 1 yr)
& $p(d_{\mathrm{L}}=200$ Mpc; 1 day)\\
(Myr$^{-1}$)&Initial LIGO & Advanced LIGO \\
\hline
292  & 0.13 & 0.86 \\
17 & 0.008 & 0.11\\
\hline\\
\end{tabular}
\end{table}

\begin{figure}
\includegraphics[scale=0.6]{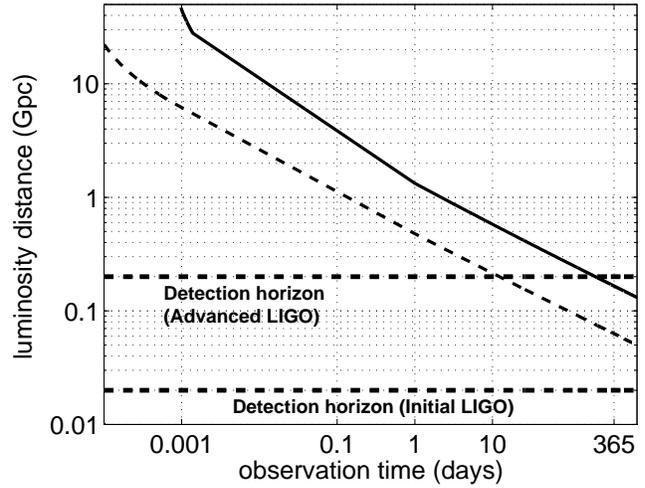} \caption{The PEH evolution for DNS mergers
as a function of observation time assuming the Galactic merger
rate limits from Kalogera et al. (2004): 292 Myr$^{-1}$ (dashed
curve) and 17 Myr$^{-1}$ (solid line), and merger rate evolution
following the instantaneous SFR. The horizontal lines show the 200
and 20 Mpc horizons, indicating the sensitivity
limits for DNS coalescence detection assuming optimal filtering
for Advanced and Initial LIGO.}\label{peh1}
\end{figure}

Table \ref{tbl-2} uses equation (\ref{prob2}) to calculate the
probability for at least one DNS merger to occur within the volume
bounded by $d_{\mathrm{L}}=20$ Mpc for one year of observation
(Initial LIGO) and within $d_{\mathrm{L}}=200$ Mpc for one day
(Advanced LIGO) using the maximum and minimum values of the
Galactic DNS merger rate $R_{\mathrm {DNS}}$ as estimated by
\cite{Kalog04}, namely 292 and 17 Myr$^{-1}$. The two
distances, 20 and 200 Mpc, correspond approximately to the
detection thresholds for DNS merger detection for Initial and
Advanced LIGO respectively. Figure \ref{peh1} plots
$d_{\mathrm{L}\epsilon}^{\mathrm{PEH}}(T)$, using the upper and
lower Galactic merger rate limits from Kalogera et al. (2004).

\section{The PEH for LIGO to detect DNS mergers}
For a DNS merger at distance $r$ from the detector, the optimal GW
signal-to-noise ratio, $\rho$, assuming matched filtering is
\citep{hughes02}:
\begin{equation}\label{snr1}
\rho=11.7
\frac{{\vert\Psi\vert}^2}{2.56}\Big(\frac{r_{\mathrm{LIGO}}
}{r}\Big)
\end{equation}
where $r_{\mathrm{LIGO}}= 20$ and 200 Mpc for Initial and Advanced
LIGO respectively and $\vert\Psi\vert^2$, the angular sensitivity
function of the detectors, averages to 2.56 over all sky and
source positions. By substituting
$d_{\mathrm{L}\epsilon}^{\mathrm{PEH}}(T)$ for $r$ in equation
(\ref{snr1}), we obtain the DNS merger probability event horizon
of the optimal signal-to-noise ratio:
\begin{equation}\label{snr2}
\rho_{\epsilon}^{\mathrm{PEH}}(T)=11.7
\frac{{\vert\Psi\vert}^2}{2.56}\Big(\frac{r_{\mathrm{LIGO}}
}{d_{\mathrm{L}\epsilon}^{\mathrm{PEH}}(T)}\Big)\;.
\end{equation}

\begin{figure}
\includegraphics[scale=0.6]{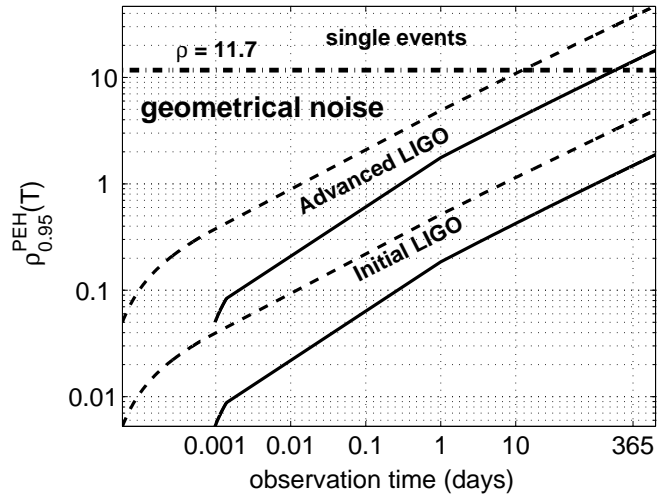} \caption{The 95\% PEH for the temporal
evolution of the signal-to-noise ratio
$\rho_{\epsilon}^{\mathrm{PEH}}(T)$ for Advanced and Initial LIGO
using the local Galactic rate limits for DNS mergers from Kalogera
et al. (2004): 292 Myr$^{-1}$ (dashed curve) and 17 Myr$^{-1}$
(solid line). For Advanced LIGO with detection threshold
$\rho>11.7$ (assuming optimal filtering), at least one DNS merger
will be detectable during 12--211 days of observation. For initial
LIGO sensitivities, DNS mergers will be detectable in 22--380 yrs
of observation and, for a $\rho \sim 1$, in 4--68 days. In the
region below $\rho=11.7$, individual mergers are unlikely to be
detected, but the cumulative signal will manifest as `geometrical
noise' in interferometric data.}\label{fig3}
\end{figure}

Assuming $\rho \sim 11.7$, corresponding to the detection
threshold for a DNS merger in Advanced LIGO for an `average'
source at 200 Mpc \citep{hughes02}, one can model the evolution of
$\rho$ as a function of duty cycle or observation time. Figure
\ref{fig3} plots $\rho_{0.95}^{\mathrm{PEH}}(T)$, showing the PEH
for at least one DNS merger to occur during observation time $T$,
with $\rho \ge \rho_{0.95}^{\mathrm{PEH}}(T)$, for the upper and
lower uncertainty limits of the local DNS rate densities
\citep{Kalog04}. For Advanced LIGO, at least one DNS merger with a
 $\rho>$ 11.7 will be detectable using a matched
filter every 12--211 days with a 95\% probability. For initial
LIGO sensitivities, events with the same $\rho$ will be detectable
at a rate of one per 22--380 yr and for $\rho \sim 1$, one every
4--68 days. Although DNS mergers with $\rho\la 11.7$ will be
present in the data, they are unlikely to be temporally resolved.

Here, we consider a signal comprised of an ensemble of chirp signals that result from the final mergers of cosmological DNSs. Even though the individual signals will not overlap, the cumulative background is stochastic because one cannot predict with certainty the exact temporal evolution of the signal. But, the GW amplitude distribution will be very different from a continuous astrophysical GW background. For  a signal that consists of a sum of many continuously emitting, randomly distributed sources, such as the GW signal from the Galactic population of white dwarf-white dwarf or white dwarf-NS binaries, the central limit theorem predicts that the GW amplitude distribution will be Gaussian. The term `confusion noise' is often used to describe the signal from the Galactic population of white dwarf-white dwarf binaries. In contrast, the GW background from DNS mergers has a non-Gaussian signature that is characterized by the spatial distribution of the sources, so we use the term geometrical noise to describe it.

\section{The PEH technique applied to simulated data}
\cite{cow02a} developed a procedure to simulate time series of the
GW signals from supernovae occurring throughout the Universe. The
method has provided a tool to probe the statistical signature of
cosmological `standard candle' GW sources that follow the evolving
star formation rate. In its simplest form, the simulation outputs
a distribution of events in redshift as a function of observation
time. Here, we use this procedure to extract the PEH signature
from a distribution of observer-source distances and observation
times. The PEH technique is a procedure that records the time and
distance of successively closer events, ($t_i$, $r_i$), that
satisfy the condition $r_{i+1}<r_i$: each event in the PEH time
series occurs at a shorter distance than the preceding one. A key
feature of the procedure is that it probes the rate of an
observer's penetration into the small probability `tail' of an
event distribution.

\cite{cow05} applied this technique to the observed gamma-ray
burst (GRB) redshift and observation-time distribution and found
that it can be used to set limits on the local rate density of
GRBs. Because it is sensitive to the distribution of events in the
small probability tail, in this case corresponding to smaller
volumes, the procedure `tracks' the temporal evolution of close
events as detected by an observer. If a time series is comprised
of more than one class of event, then the PEH technique can
identify these distributions based on the distribution of PEH
data. This was shown by \cite{cow05}, where the PEH model fitted
to the data indicated that the anomalously close GRB 980245
probably does not belong to the `classical' GRB population.

As a further proof of concept, we apply the PEH technique to a
distribution of GW amplitudes from a simulated distribution of DNS
mergers throughout the Universe. We use a DNS dimensionless characteristic GW
amplitude $h_{c} = 4 {\times} 10^{-21}$ at 20 Mpc \citep{Schutz91}
and a non-redshifted characteristic frequency $f_{c} = 200$ Hz.
The results are plotted in figure \ref{algor}. The 5\% and 95\%
PEH curves are shown along with the output from the PEH simulation
shown as data points. The 5\% PEH curve is interpreted as the
`null PEH', as there is a 95\% probability that no events will be
observed at an amplitude above this threshold.

In order to develop the PEH concept into a technique for GW data
analysis, we need to incorporate a method for treating detector
noise. For `standard candle' sources, this can be done by working
in terms of progressively increasing amplitudes rather than
progressively decreasing redshifts, as shown in figure 5. Noise
transients of varying amplitudes mimic sources at varying
redshifts, enabling the introduction of a `noise PEH'. Here we use
a `Gaussian-noise PEH'; this is based on a standard zero-mean
Gaussian distribution with standard deviation $s_d =
h_{rms}\sqrt{f_s/2f_c}$, modelled on idealized interferometric
detector noise with $h_{rms}$ denoting the root-mean-square
amplitude of the noise and $f_s$ the sampling frequency--see
\cite{Arn03}. Figure 5 shows the 5\% and 95\% Gaussian-noise PEHs.

\begin{figure}
\includegraphics[scale=0.65]{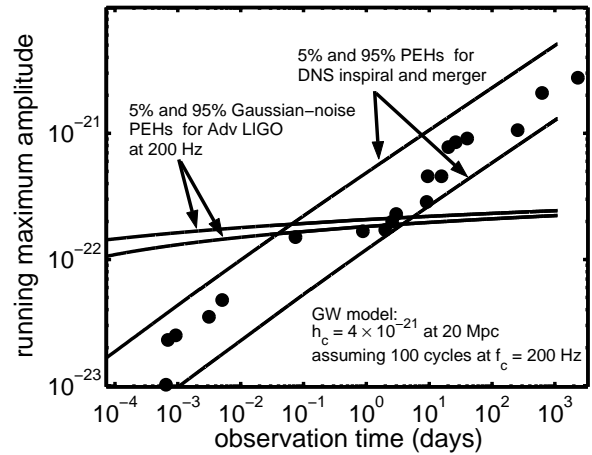} \caption{ The PEH technique applied to a
distribution of GW amplitudes from simulated DNS mergers throughout the
Universe assuming the upper Galactic rate limit of 292 Myr$^{-1}$
from Kalogera et al. (2004). We use a DNS characteristic GW
amplitude $h_{c} = 4 {\times} 10^{-21}$ at 20 Mpc (Schutz 1991)
and a non-redshifted characteristic frequency $f_{c} = 200$ Hz.
The 5\% (or 95\% probability of no events occurring) and 95\% PEH
curves for DNS mergers are shown, along with the output of the PEH
simulation applied to a distribution of characteristic GW
amplitudes modelled on DNS mergers, shown as filled circles. Also
shown are the 5\% and 95\% `Gaussian-noise PEHs', modelled on
idealized interferometric detector noise--see text for details.
The figure shows that the DNS merger PEHs evolve much faster than
the Gaussian-noise PEHs.}
\end{figure}\label{algor}

It is clear from the small gradient of the Gaussian-noise PEH
curves that the temporal evolution of transients, originating from
the tail of the distribution is slow. The striking feature of this
figure is the gradient of the PEH curves for the source amplitude
distribution: it is much bigger than for the Gaussian-noise case,
because of the distance-amplitude relation. We define this type of
PEH data as `geometrical', because it is linked to the spatial
distribution of sources. 

We note that although we have incorporated SFR evolution and a cosmological model
to calculate the DNS merger rate at high $z$, the PEH will be most sensitive to events that are comparable to or greater than the detection threshold
of the detector. For Advanced LIGO, this will be of order several hundred Mpc ($z\approx 0.05$), 
implying that the PEH will be dominated by sources in the local
Universe where cosmological effects are minor. 

Given that a geometrical signature is present in a time series,
one can fit a PEH model to the data, with the local rate density
as a free parameter. In the presence of noise, this signature will
be modified, but if the detector noise is well characterized, a
fitting function constructed from the sum of both noise and
geometrical distributions can be used. For small signal-noise
ratios, the noise will dominate the PEH data for short observation
times. But as observation time increases, the geometrical
signature will grow at a faster rate than the noise. It is
possible that the GW signal from DNS mergers can be identified
before that of a single merger, occurring above the detector
threshold, as shown in figure \ref{fig3}. We propose that there
exists an intermediate detection regime, with pre-filtered signal-noise-ratio
of less than unity, where the DNS merger rate is high enough that the
geometrical signature could be identified in interferometer data.
This type of search provides an opportunity to set upper limits on
the DNS merger rate.

\section{Discussion}

The cumulative signal from transient GW sources at cosmological
distances is commonly described as a stochastic background because
of the temporal randomness of the individual events. But we have
shown here that, for a cosmic population of DNS mergers, the GW
cumulative signal will be dominated by the spatial distribution
and temporal evolution of the sources. We use the term
`geometrical' noise, because the GW amplitude distribution
composed of DNS mergers is dominated by the spatial distribution
of the sources.

Interestingly, geometrical noise from cosmic DNS mergers (and
similar transient GW sources) is linked to the sensitivity of the
detector and the cumulative event rate. If a GW detector could
probe DNS mergers to distances corresponding to the time when they
first occurred in the Universe, the detector could potentially
resolve all events and there would be no geometrical noise from
unresolved signals. However, if the cumulative rate measured in
our frame were to be so high that the individual chirps would
overlap, corresponding to a signal of duty cycle greater than
unity, then the resulting signals would manifest as geometrical
noise. This is unlikely for DNS mergers: the individual events are
not expected to overlap but to be nonetheless not individually
detectable because of the small signal-to-noise ratio.

We highlight the concept that the detectability of a GW stochastic
background from DNS mergers throughout the Universe and of
individual DNS mergers is linked to the observer via the
instrument duty cycle, the cumulative event rate and the
sensitivity of the detector. The PEH model is applied above to the
detectability of DNS mergers with Advanced and Initial LIGO; it
shows the temporal evolution of the GW signal from many faint
unresolved events converging to detectable single events of high
signal-to-noise ratio. Detection strategies based on the PEH model
could utilize the geometrical signature of the temporal evolution
of the chirp GW amplitude distribution. We plan to investigate the
PEH technique further using simulated GW background and
non-Gaussian noise data.

The PEH method could also be used to set constraints on the black hole-black hole (BH-BH) merger rate, which is presently highly uncertain. Although occurring at a lower intrinsic rate than DNS mergers, the GW emission from BH-BH mergers is expected to be significantly higher, implying an enhanced detection rate because of the larger detection volume. Estimates range from 30--4000 yr$^{-1}$ assuming a BH mass of 10 M$_{\odot}$ and a detection horizon of $z=0.4$ \citep{Thorne02}  for Advanced LIGO sensitivity. We plan to use the BH-BH mass and merger rate distributions to determine if a PEH model could be developed and applied to interferometer data to set limits on the rate and mass distributions.

Although we have shown how the cumulative GW signal from DNS
mergers should manifest in a time series, we have not yet tested
how robust the above procedure is to the non-stationarity of the
data. For instance, if a time series is plagued with non-Gaussian
transients, then the effectiveness of the fitting procedure will
be reduced, unless a good model for the transients is available.

We assume optimal (matched) filtering can be applied to the candidate transients, but the
filters that will be most effective will not be exactly matched to individual events. The `norm' and `mean' filters, although not optimal, should perform at a significant fraction of a matched filter for transients where the signal is not completely characterized \citep{Arn03}. We emphasize that using the PEH in a signal processing context is based on fitting to outliers using a model distribution and not `detecting'  individual events.  Nonetheless, it is important to filter the data to select as many transients that are candidates for DNS mergers as possible. As a test of the robustness of the procedure using different filters, we plan to inject
simulated cosmologically distributed GW transients into
non-Gaussian noise.  

We note that the simulated GW amplitude distribution will be modified
by the detector antenna pattern, which we have not included in our simulations, where we used the mean value.  In applying a PEH model to real data, account must be taken of the modulation and distortion of the signal by detector characteristics and by both detector and environmental noise. We plan to extend the current simulations by incorporating such effects as part of a program to test the sensitivity of the PEH fitting procedure.

\section*{Acknowledgments}
D. M. Coward is supported by an Australian Research Council
fellowship and M. Lilley is supported from the Australian Research
Council grant DP0346344. This research is part of the research
program of the Australian Consortium for Interferometric
Gravitational Astronomy. We thank Dr Paul Saulson (LIGO) and the referee
for several helpful comments and suggestions.

\label{lastpage}

\end{document}